\pdfoutput=1
\documentclass{article}
\usepackage[preprint]{spconf}
\usepackage{amsmath,amssymb,graphicx,bm,booktabs,multirow}
\usepackage{array,ragged2e}
\newcolumntype{L}[1]{>{\RaggedRight\arraybackslash}p{#1}}
\usepackage[hidelinks]{hyperref}
\usepackage{tikz}
\usepackage{xcolor}
\usetikzlibrary{positioning,arrows.meta,fit,backgrounds,calc,shapes.geometric}
\usepackage{pgfplots}
\usepackage{float}
\pgfplotsset{compat=1.18}

\definecolor{cteal}{RGB}{0,160,170}
\definecolor{colive}{RGB}{124,172,72}
\definecolor{cslate}{RGB}{92,126,186}
\definecolor{cterra}{RGB}{219,108,55}
\definecolor{ctan}{RGB}{182,152,92}
\definecolor{cnavy}{RGB}{24,62,118}
\definecolor{cgold}{RGB}{238,168,40}
\definecolor{mnavy}{RGB}{58,82,134}
\definecolor{mblue}{RGB}{122,152,196}
\definecolor{mred}{RGB}{188,98,86}
\definecolor{mgreen}{RGB}{138,158,118}
\definecolor{mpurple}{RGB}{146,128,162}
\definecolor{msand}{RGB}{192,172,132}
\definecolor{fblue}{RGB}{31,119,180}
\definecolor{forange}{RGB}{217,83,25}
\definecolor{fgold}{RGB}{237,177,32}
\definecolor{fgreen}{RGB}{119,172,48}
\definecolor{fpurple}{RGB}{126,47,142}
\definecolor{fcyan}{RGB}{77,190,238}

\newcommand{\bh}{\mathbf{h}}
\newcommand{\bH}{\mathbf{H}}
\newcommand{\by}{\mathbf{y}}
\newcommand{\bu}{\mathbf{u}}

\newcommand{\bp}{\mathbf{p}^{\mathrm{UE}}}
\newcommand{\hbp}{\hat{\mathbf{p}}^{\mathrm{UE}}}
\newcommand{\bb}{\mathbf{p}^{\mathrm{BS}}}
\newcommand{\bR}{\mathbf{R}}
\newcommand{\bI}{\mathbf{I}}
\newcommand{\bV}{\mathsf{V}}
\newcommand{\cA}{\mathcal{A}}
\newcommand{\cD}{\mathcal{D}}
\newcommand{\cP}{\mathcal{P}}
\newcommand{\E}{\mathbb{E}}
\newcommand{\Np}{N_{\mathrm{p}}}

\title{Geometry-Informed Neural Network for Pilot-Limited Channel Estimation via Sensing from Uplink Traffic}

\name{Wenyu Huang, Alireza Javid, and Nuria Gonz\'alez-Prelcic}
\address{University of California San Diego, USA}

\makeatletter
\renewcommand\section{\@startsection{section}{1}{\z@}%
  {-1.95ex \@plus -0.5ex \@minus -.2ex}%
  {1.0ex \@plus .2ex}%
  {\normalfont\normalsize\bfseries}}
\renewcommand\subsection{\@startsection{subsection}{2}{\z@}%
  {-1.8ex \@plus -0.5ex \@minus -.2ex}%
  {0.7ex \@plus .2ex}%
  {\normalfont\normalsize\bfseries}}
\newlength{\tabcapabove}\newlength{\tabcapbelow}
\def\spconf@tabletype{table}
\long\def\@makecaption#1#2{%
  \ifx\@captype\spconf@tabletype\vskip\tabcapabove\else\vskip\abovecaptionskip\fi
  \setbox\@tempboxa\hbox{#1. #2}%
  \ifdim \wd\@tempboxa >\hsize #1. #2\par \else
    \hbox to\hsize{\hfil\box\@tempboxa\hfil}\fi
  \ifx\@captype\spconf@tabletype\vskip\tabcapbelow\fi}
\let\spconf@small\small
\renewcommand\small{\spconf@small
  \abovedisplayskip 4\p@ \@plus 2\p@ \@minus 2\p@
  \abovedisplayshortskip \z@ \@plus 2\p@
  \belowdisplayskip \abovedisplayskip
  \belowdisplayshortskip 3\p@ \@plus 2\p@ \@minus 2\p@}
\makeatother

\toappear{\parbox[b]{\textwidth}{\centering\footnotesize This work has been submitted to the IEEE for possible publication. Copyright may be transferred without notice, after which this version may no longer be accessible.}}

\begin{document}
\raggedbottom
\ninept
\renewcommand{\baselinestretch}{0.88}\normalsize
\maketitle
\begin{abstract}

Reliable wireless systems depend on accurate uplink channel estimation for precoding, scheduling, and link adaptation at the base station.
In the upper mid-band, wide bandwidths and large antenna arrays increase the channel dimension and make pilot-efficient estimation particularly challenging.
While uplink data is available as every user transmits it, sounding resources are shared across all users in the cell, leaving only a few pilots for each user.
Thus, channel estimation heavily depends on prior information. Existing approaches obtain such priors in different ways, each with limitations. Ray tracing requires a site model, channel knowledge maps rely on measurements at or near the locations they represent, and learned channel distributions do not explicitly capture propagation geometry.
We therefore propose TRaffic-Aided Channel Estimation (TRACE), which builds the prior from uplink traffic without additional sounding.
TRACE learns a geometry-aware channel prior from routine multi-UE uplink traffic and combines it with the available pilots for channel estimation. In a ray-traced street canyon at \(15\,\mathrm{GHz}\), TRACE achieves \(-7.48\,\mathrm{dB}\) NMSE with 8 pilots, outperforming three state-of-the-art priors and two geometric ablations by more than \(3\,\mathrm{dB}\), while remaining effective with only 2 pilots.
Within a deployment, TRACE remains robust across SNRs, carrier frequencies, and reduced mapping traffic without retraining.

\end{abstract}


\section{Introduction}
\label{sec:intro}

Reliable wireless systems depend on accurate uplink channel estimation to support precoding, scheduling, and link adaptation at the base station (BS).
In the upper mid-band, wide bandwidths and large antenna arrays increase the channel dimension, making pilot-efficient estimation particularly challenging~\cite{javid2026rss}.
The challenge is especially pronounced on the uplink since every user routinely transmits data on the physical uplink shared channel (PUSCH), whereas sounding reference signal (SRS) resources must be shared across all users in the cell.
Consequently, the BS observes abundant uplink traffic but may have only a few pilots to estimate each user's channel.

With few pilots, many channels fit the observations, making the estimate strongly dependent on the prior and how it is obtained.
Foundation models use past channels to predict future ones~\cite{zhou2025pfm}, but require user-specific history that is not available to newly arrived user equipment (UE).
Site-specific methods, including neural radio fields~\cite{lu2024newrf,umer2025ngrf}, point-cloud methods~\cite{xue2026pctgs}, and learned simulators~\cite{hehn2025wigatr}, require measurements or explicit 3D geometry collected in advance.
Ray tracing on an uncalibrated digital twin can predict channel statistics from few measurements~\cite{abouamer2025}, but still requires a 3D model of the site.
Channel knowledge maps (CKMs), including IC-CKM~\cite{icckm2025}, learn location-dependent channel information from received signals~\cite{wu2024ckm,jiang2025dynamicckm,cao2025dmd}, but rely on interpolation or prediction at locations without direct measurements.
Some methods estimate geometry during service.
Bistatic receivers turn their own path estimates into a covariance~\cite{wang2026bistatic}, radio simultaneous localization and mapping (SLAM) jointly estimates the user and environment map~\cite{deutschmann2025geometry}, and Gaussian scene models learn a delay and beam prior along scheduled tracks~\cite{geogs2026}.
However, these methods need online sensing or trajectory-specific structure, and infer geometry separately for each UE.
Score-based generative channel models~\cite{zhou2025diffusion,arvinte2023score} and variational autoencoders for covariance-based MMSE estimation~\cite{Baur2024} avoid explicit geometry, but learn black-box priors directly from channel samples.
Environmental priors can also be built from RSS maps~\cite{javid2026rss} or path-level radio maps~\cite{wang2026charm}, but these require offline site-specific maps, obtained from measurements, ray tracing, or a digital twin.
Taken together, existing methods rely on priors from user history, site-specific information, additional measurements, or black-box learning, and infer geometry online for each user separately.

We instead seek a prior that requires neither a site survey nor extra pilots, captures the cell geometry, and can be adapted to a new UE from only a coarse position. We propose TRaffic-Aided Channel Estimation (TRACE), which uses routine PUSCH traffic as a sensing source to build the channel prior without additional signaling, following the integrated sensing and communication (ISAC) idea of reusing communication signals for sensing.
After timing and frequency synchronization, PUSCH bursts provide noisy angle, delay, and Doppler detections, but detections collected across UEs reflect the same propagation environment. TRACE learns this from historical traffic and resolves the instantaneous channel with the current pilots.
In deployment, training targets can be obtained from successfully decoded and remodulated PUSCH data, so no survey is needed.

Our contributions are as follows.
First, we introduce a traffic-derived spatial prior that accumulates noisy multi-UE PUSCH detections on a 3D voxel grid as shared propagation evidence, without site surveys or additional sounding.
To our knowledge, this is the first channel prior built by aggregating passive multi-UE uplink detections to learn shared propagation geometry.
Second, we develop a query-conditioned 3D U-Net that converts this shared evidence into uncertain, UE-specific path atoms, which define a covariance prior for linear minimum mean squared error (LMMSE) estimation with the current pilots. The network is trained end-to-end through this fixed estimator.
Third, in a route-wise evaluation protocol over unseen positions, TRACE outperforms three state-of-the-art (SOTA) priors and two geometric ablations while remaining robust across pilot budgets, SNRs, carrier frequencies, and reduced historical traffic.
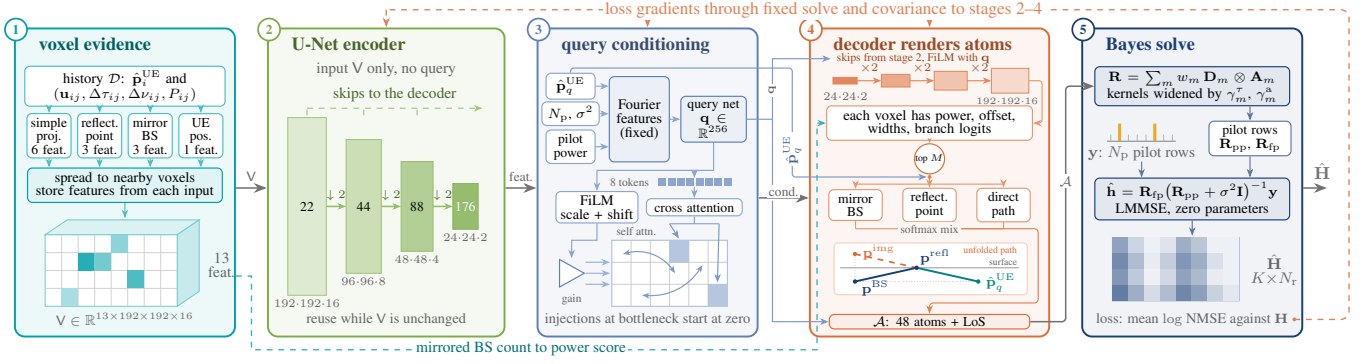
\begin{figure*}[t]
\renewcommand{\hbp}{\hat{\mathbf{p}}^{\mathrm{UE}}}
\centering
\resizebox{\textwidth}{!}{%
\begin{tikzpicture}[
  font=\scriptsize,
  panel/.style={rounded corners=1.8mm, line width=1pt},
  ptitle/.style={font=\fontsize{9.5}{10.5}\selectfont\bfseries, anchor=north west},
  pcap/.style={font=\fontsize{7.5}{8}\selectfont, text=black!60, anchor=south, align=center},
  badge/.style={circle, fill=white, line width=0.9pt, inner sep=0.6pt, font=\scriptsize\bfseries, minimum size=4mm},
  blk/.style={draw=#1, fill=white, rounded corners=0.9mm, line width=0.6pt, font=\scriptsize, align=center, inner sep=2pt},
  tblk/.style={draw=#1, fill=#1!12, rounded corners=0.9mm, line width=0.6pt, font=\scriptsize, align=center, inner sep=2pt},
  op/.style={circle, draw=#1, fill=white, line width=0.6pt, font=\scriptsize, inner sep=0.6pt},
  arr/.style={-{Stealth[length=2.4mm]}, thick, black!55},
  sarr/.style={-{Stealth[length=1.6mm]}, line width=0.6pt, #1},
  bus/.style={line width=0.6pt, #1},
  flow/.style={font=\scriptsize, text=black!60, anchor=south},
  note/.style={font=\fontsize{8}{8.5}\selectfont, text=black!60, align=center},
  tl/.style={font=\fontsize{6}{6.5}\selectfont, text=black!60, align=center}]
\renewcommand{\hbp}{\hat{\mathbf{p}}^{\mathrm{UE}}}
\begin{scope}[shift={(0,0)}]
\draw[panel, draw=cteal, fill=cteal!7] (0,0) rectangle (4.0,5.60);
\node[ptitle, text=cteal!60!black] at (0.32,5.55) {voxel evidence};
\node[blk=cteal!70, text width=3.5cm, font=\fontsize{7.5}{8}\selectfont] (hist) at (2.0,4.55)
  {history $\cD$: $\hbp_i$ and $(\bu_{ij},\Delta\tau_{ij},\Delta\nu_{ij},P_{ij})$};
\node[blk=cteal!70, text width=0.66cm] (b1) at (0.62,3.65) {simple\\proj.\\6 feat.};
\node[blk=cteal!70, text width=0.66cm] (b2) at (1.54,3.65) {reflect.\\point\\3 feat.};
\node[blk=cteal!70, text width=0.66cm] (b3) at (2.46,3.65) {mirror\\BS\\3 feat.};
\node[blk=cteal!70, text width=0.66cm] (b4) at (3.38,3.65) {UE\\pos.\\1 feat.};
\foreach \n in {b1,b2,b3,b4} \draw[sarr=cteal!70] (hist.south -| \n) -- (\n.north);
\node[tblk=cteal, text width=3.4cm, font=\fontsize{7.5}{8}\selectfont] (splat) at (2.0,2.75)
  {spread to nearby voxels\\store features from each input};
\foreach \n in {b1,b2,b3,b4} \draw[sarr=cteal!70] (\n.south) -- (splat.north -| \n);
\fill[cteal!14, draw=cteal!55] (0.55,1.85) -- (2.95,1.85) -- (3.35,2.12) -- (0.95,2.12) -- cycle;
\fill[cteal!22, draw=cteal!55] (2.95,0.55) -- (3.35,0.82) -- (3.35,2.12) -- (2.95,1.85) -- cycle;
\fill[white, draw=cteal!55] (0.55,0.55) rectangle (2.95,1.85);
\foreach \i in {1,...,7} \draw[gray!35] (0.55+\i*0.3,0.55) -- (0.55+\i*0.3,1.85);
\foreach \j in {1,...,3} \draw[gray!35] (0.55,0.55+\j*0.325) -- (2.95,0.55+\j*0.325);
\fill[cteal!85] (1.15,1.20) rectangle (1.45,1.525);
\fill[cteal!55] (1.45,1.20) rectangle (1.75,1.525);
\fill[cteal!70] (2.05,0.875) rectangle (2.35,1.20);
\fill[cteal!35] (0.85,0.55) rectangle (1.15,0.875);
\fill[cteal!45] (1.75,1.525) rectangle (2.05,1.85);
\draw[sarr=cteal!70] (splat.south) -- (2.0,2.15);
\node[note, anchor=west] at (3.4,1.3) {13\\feat.};
\node[pcap] at (2.0,0.08) {$\bV\in\mathbb R^{13\times192\times192\times16}$};
\end{scope}
\begin{scope}[shift={(4.6,0)}]
\draw[panel, draw=colive, fill=colive!7] (0,0) rectangle (4.3,5.60);
\node[ptitle, text=colive!60!black] at (0.32,5.55) {U-Net encoder};
\node[note] at (2.15,4.86) {input $\bV$ only, no query};
\draw[colive!70, dashed, line width=0.6pt, -{Stealth[length=1.6mm]}] (0.71,4.15) -- (3.9,4.15)
  node[pos=0.5, above=0.2pt, font=\footnotesize, text=colive!60!black] {skips to the decoder};
\foreach \x in {0.71,1.74,2.71} \draw[colive!70, dashed, line width=0.6pt] (\x,4.02) -- (\x,4.15);
\fill[colive!25, draw=colive!70] (0.36,0.75) rectangle (1.06,3.98);
\fill[colive!40, draw=colive!75] (1.42,1.15) rectangle (2.06,3.58);
\fill[colive!58, draw=colive!80] (2.42,1.55) rectangle (3.00,3.18);
\fill[colive!80, draw=colive!90] (3.36,1.95) rectangle (3.82,2.78);
\node[font=\scriptsize, text=black] at (0.71,2.365) {22};
\node[font=\scriptsize, text=black] at (1.74,2.365) {44};
\node[font=\scriptsize, text=black] at (2.71,2.365) {88};
\node[font=\scriptsize, text=white] at (3.59,2.365) {176};
\foreach \x/\y in {1.06/1.42, 2.06/2.42, 3.00/3.36} \draw[sarr=colive!80, thick] (\x,2.365) -- (\y,2.365);
\node[tl, text=colive!60!black] at (1.24,2.58) {$\downarrow2$};
\node[tl, text=colive!60!black] at (2.24,2.58) {$\downarrow2$};
\node[tl, text=colive!60!black] at (3.18,2.58) {$\downarrow2$};
\node[tl] at (0.71,0.6) {$192{\cdot}192{\cdot}16$};
\node[tl] at (1.74,1.0) {$96{\cdot}96{\cdot}8$};
\node[tl] at (2.71,1.4) {$48{\cdot}48{\cdot}4$};
\node[tl] at (3.59,1.8) {$24{\cdot}24{\cdot}2$};
\node[pcap] at (2.15,0.08) {reuse while $\bV$ is unchanged};
\end{scope}
\begin{scope}[shift={(9.5,0)}]
\draw[panel, draw=cslate, fill=cslate!7] (0,0) rectangle (4.0,5.60);
\node[ptitle, text=cslate!60!black] at (0.32,5.55) {query conditioning};
\node[blk=cslate!70, text width=0.75cm] (in1) at (0.6,4.55) {$\hbp_q$};
\node[blk=cslate!70, text width=0.9cm] (in2) at (0.6,4.05) {$\Np$, $\sigma^2$};
\node[blk=cslate!70, text width=0.75cm] (in3) at (0.6,3.45) {pilot\\power};
\node[tblk=cslate, text width=0.95cm, minimum height=1.55cm, font=\fontsize{7.5}{8}\selectfont] (ff) at (1.85,3.95) {Fourier\\features\\(fixed)};
\foreach \n in {in1,in2,in3} \draw[sarr=cslate!70] (\n.east) -- (ff.west |- \n);
\node[tblk=cslate, text width=1.05cm] (mlp) at (3.2,3.95) {query net\\$\mathbf q\in\mathbb R^{256}$};
\draw[sarr=cslate!70] (ff.east) -- (mlp.west);
\node[blk=cslate!70, text width=1.35cm, font=\fontsize{7.5}{8}\selectfont] (film) at (1.08,2.35) {FiLM\\scale + shift};
\node[blk=cslate!70, text width=1.6cm] (xat) at (2.85,2.35) {cross attention};
\draw[bus=cslate!70] (mlp.south) -- (3.2,3.02) -- (1.08,3.02);
\draw[sarr=cslate!70] (1.08,3.02) -- (film.north);
\draw[sarr=cslate!70] (2.85,3.02) -- (2.85,2.88);
\foreach \k in {0,...,7} \fill[cslate!70] (2.2+\k*0.17,2.74) rectangle (2.33+\k*0.17,2.87);
\draw[sarr=cslate!70] (2.85,2.72) -- (xat.north);
\draw[bus=cslate!70] (xat.south) -- (2.85,2.05) (2.4,2.05) -- (3.3,2.05);
\node[tl, anchor=east] at (2.14,2.805) {8 tokens};
\fill[white, draw=cslate!55] (1.35,0.55) rectangle (3.45,1.75);
\foreach \i in {1,...,5} \draw[gray!40] (1.35+\i*0.35,0.55) -- (1.35+\i*0.35,1.75);
\foreach \j in {1,2} \draw[gray!40] (1.35,0.55+\j*0.4) -- (3.45,0.55+\j*0.4);
\fill[cslate!45] (2.4,1.35) rectangle (2.75,1.75);
\fill[cslate!45] (3.1,0.55) rectangle (3.45,0.95);
\draw[sarr=cslate!80] (2.4,2.05) -- (2.57,1.78);
\draw[sarr=cslate!80] (3.3,2.05) -- (3.27,0.98);
\draw[sarr=cslate!80, {Stealth[length=1.4mm]}-{Stealth[length=1.4mm]}] (1.55,1.5) to[bend left=35] (2.5,1.15);
\draw[sarr=cslate!80, {Stealth[length=1.4mm]}-{Stealth[length=1.4mm]}] (1.9,0.75) to[bend right=30] (3.0,1.3);
\node[tl, anchor=west] at (1.32,1.92) {self attn.};
\draw[cslate, fill=white, line width=0.6pt] (0.4,1.4) -- (0.4,0.9) -- (0.85,1.15) -- cycle;
\draw[sarr=cslate!70] (film.south) -- (1.08,1.75) -- (0.62,1.75) -- (0.62,1.42);
\foreach \yy in {1.35,1.15,0.95} \draw[sarr=cslate!60] (0.87,\yy) -- (1.32,\yy);
\node[tl] at (0.62,0.68) {gain};
\node[pcap] at (2.0,0.08) {injections at bottleneck start at zero};
\end{scope}
\begin{scope}[shift={(14.45,0)}]
\draw[panel, draw=cterra, fill=cterra!7] (0,0) rectangle (4.35,5.60);
\node[ptitle, text=cterra!70!black] at (0.32,5.55) {decoder renders atoms};
\node[tl, text=cterra!70!black, anchor=west] (dec) at (0.38,5.05)
  {skips from stage 2, FiLM with $\mathbf q$};
\foreach \xa/\xb/\h/\lvl in {0.42/0.86/0.12/80, 1.30/1.82/0.23/58, 2.26/2.86/0.34/40, 3.30/3.98/0.45/25}
  \fill[cterra!\lvl, draw=cterra!80] (\xa,{4.65-\h/2}) rectangle (\xb,{4.65+\h/2});
\foreach \x/\y in {0.86/1.30, 1.82/2.26, 2.86/3.30}
  \draw[sarr=cterra!80] (\x,4.65) -- (\y,4.65);
\foreach \x in {1.08,2.04,3.08}
  \node[tl, text=cterra!70!black] at (\x,4.88) {$\times2$};
\node[tl] at (0.65,4.36) {$24{\cdot}24{\cdot}2$};
\node[tl, font=\tiny] at (3.59,4.29) {$192{\cdot}192{\cdot}16$};
\node[blk=cterra!70, text width=3.6cm, font=\fontsize{7.5}{8}\selectfont] (nine) at (2.175,3.88)
  {each voxel has power, offset,\\widths, branch logits};
\draw[sarr=cterra!70, rounded corners=0.6mm] (3.98,4.65) -- (4.22,4.65) -- (4.22,0 |- nine.east) -- (nine.east);
\node[op=cterra, font=\fontsize{5}{5.5}\selectfont, inner sep=0.1pt, minimum size=4.8mm] (topm) at (2.175,3.23) {top $M$};
\draw[-{Stealth[length=1.0mm]}, line width=0.6pt, cterra!80] (nine.south) -- (topm.north);
\coordinate (branchin) at (0.83,2.77);
\coordinate (geomjoin) at (2.175,2.90);
\fill[cterra!85] (geomjoin) circle (0.035);
\draw[bus=cterra!80] (topm.south) -- (2.175,2.77);
\draw[bus=cterra!70] (0.83,2.77) -- (3.52,2.77);
\node[blk=cterra!70, text width=0.83cm] (va) at (0.83,2.39) {mirror\\BS};
\node[blk=cterra!70, text width=0.83cm] (ps) at (2.175,2.39) {reflect.\\point};
\node[blk=cterra!70, text width=0.83cm] (los) at (3.52,2.39) {direct\\path};
\foreach \n in {va,ps,los}
  \draw[sarr=cterra!70] (\n.north |- branchin) -- (\n.north);
\foreach \n in {va,ps,los}
  \draw[bus=cterra!70] (\n.south) -- (\n.south |- 0,1.96);
\draw[bus=cterra!70] (0.83,1.96) -- (3.52,1.96);
\node[tl, fill=cterra!7, inner sep=1.2pt] (mix) at (2.175,1.96) {softmax mix};
\node[tblk=cterra, text width=3.55cm, inner sep=1.3pt] (atoms) at (2.175,0.27)
  {$\cA$: 48 atoms + LoS};
\draw[sarr=cterra!85, rounded corners=0.7mm]
  (mix.east) -- (4.15,1.96) -- (4.15,0.61) -- (2.175,0.61) -- (atoms.north);
\fill[white, draw=cterra!35, rounded corners=0.7mm] (0.44,0.76) rectangle (3.95,1.70);
\coordinate (gbs) at (0.82,1.00);
\coordinate (gimg) at (0.82,1.50);
\coordinate (grefl) at (1.94,1.25);
\coordinate (gue) at (3.06,1.00);
\draw[black!35, line width=0.6pt] (0.55,1.25) -- (3.80,1.25);
\draw[black!25, densely dotted] (gimg) -- (gbs) -- (gue);
\draw[cterra, dashed, line width=0.9pt] (gimg) -- (grefl);
\draw[cnavy, line width=1pt] (gbs) -- (grefl);
\draw[cteal!80!black, line width=1pt] (grefl) -- (gue);
\fill[cnavy] (gbs) circle (0.04);
\fill[cterra] (gimg) circle (0.04);
\fill[cnavy] (grefl) circle (0.04);
\fill[cteal!80!black] (gue) circle (0.04);
\node[font=\scriptsize, text=cterra, anchor=west, inner sep=0.5pt] at (0.94,1.53) {$\mathbf p^{\mathrm{img}}$};
\node[font=\scriptsize, text=cnavy, anchor=south west, inner sep=0.5pt] at (1.98,1.29) {$\mathbf p^{\mathrm{refl}}$};
\node[font=\scriptsize, text=cnavy, anchor=west, inner sep=0.5pt] at (0.92,0.88) {$\bb$};
\node[font=\scriptsize, text=cteal!80!black, anchor=west, inner sep=0.5pt] at (3.18,0.98) {$\hbp_q$};
\node[tl, font=\tiny, anchor=south east, inner sep=0.5pt] at (3.79,1.27) {surface};
\node[tl, font=\tiny, text=cterra, anchor=east, inner sep=0.5pt] at (3.79,1.56) {unfolded path};
\end{scope}
\begin{scope}[shift={(19.4,0)}]
\draw[panel, draw=cnavy, fill=cnavy!6] (0,0) rectangle (4.0,5.60);
\node[ptitle, text=cnavy] at (0.32,5.55) {Bayes solve};
\node[tblk=cnavy, text width=3.4cm, font=\fontsize{7.5}{8}\selectfont] (cov) at (2.0,4.55) {$\bR=\sum_m w_m\,\mathbf D_m\otimes\mathbf A_m$\\kernels widened by $\gamma^\tau_m$, $\gamma^{\mathrm a}_m$};
\draw[black!50] (0.45,3.55) -- (1.7,3.55);
\foreach \x in {0.52,0.68,0.84,1.0,1.16,1.32,1.48,1.64} \draw[gray!60] (\x,3.55) -- (\x,3.67);
\draw[cgold, line width=1.6pt] (0.68,3.55) -- (0.68,3.87);
\draw[cgold, line width=1.6pt] (1.32,3.55) -- (1.32,3.87);
\node[note] at (1.08,3.33) {$\by$: $\Np$ pilot rows};
\node[blk=cnavy!70, text width=1.35cm] (sel) at (3.05,3.6) {pilot rows\\$\bR_{\mathrm{pp}}$, $\bR_{\mathrm{fp}}$};
\draw[sarr=cnavy!70] (cov.south -| sel) -- (sel.north);
\node[tblk=cnavy, text width=3.4cm, font=\fontsize{7.5}{8}\selectfont] (lmmse) at (2.0,2.5) {$\hat\bh=\bR_{\mathrm{fp}}\big(\bR_{\mathrm{pp}}+\sigma^2\bI\big)^{-1}\by$\\LMMSE, zero parameters};
\draw[sarr=cnavy!70] (sel.south) -- (lmmse.north -| sel);
\draw[sarr=cnavy!70] (1.08,3.18) -- (lmmse.north -| 1.08,3.18);
\fill[white, draw=cnavy!55] (0.55,0.65) rectangle (2.85,1.8);
\foreach [count=\i from 0] \c in {68,42,18,34,62,50,24,12}{
  \foreach [count=\j from 0] \r in {0,14,8,20}{
    \pgfmathtruncatemacro{\lvl}{max(8,\c-\r)}
    \fill[cnavy!\lvl] (0.55+\i*0.2875,0.65+\j*0.2875) rectangle (0.55+\i*0.2875+0.2875,0.65+\j*0.2875+0.2875);
  }}
\draw[sarr=cnavy!70] (lmmse.south) -- (2.0,1.83);
\node[note, anchor=west] at (2.95,1.22) {$\hat\bH$\\$K{\times}N_{\mathrm r}$};
\node[pcap] (trainloss) at (2.0,0.08) {loss: mean $\log$ NMSE against $\bH$};
\end{scope}
\draw[sarr=cslate!80] (mlp.east) -- (13.77,3.95) -- (13.77,0 |- dec.west) -- (dec.west);
\draw[sarr=cslate!80] (13.77,3.95) -- (13.77,0 |- atoms.west) -- (atoms.west);
\node[tl, rotate=90, fill=white, inner sep=1pt] at (13.77,4.48) {$\mathbf q$};
\draw[sarr=cslate!80, rounded corners=0.6mm]
  (in1.north) -- (in1.north |- 0,4.88) -- (14.13,4.88) -- (14.13,0 |- geomjoin) -- ([xshift=-0.5mm]geomjoin);
\node[font=\scriptsize, text=cslate!80!black, rotate=90, fill=white, inner sep=1.2pt]
  at (14.13,3.40) {$\hbp_q$};
\draw[cteal!80, dashed, line width=0.7pt, -{Stealth[length=1.6mm]}]
  (4.02,1.1) -- (4.3,1.1) -- (4.3,-0.24) -- (14.58,-0.24) -- (14.58,0 |- nine.west) -- (nine.west);
\node[font=\footnotesize, text=cteal!70!black, fill=white, inner sep=1pt]
  at (9.2,-0.22) {mirrored BS count to power score};
\node[badge, draw=cteal, text=cteal] at (0.02,5.58) {1};
\node[badge, draw=colive, text=colive] at (4.62,5.58) {2};
\node[badge, draw=cslate, text=cslate] at (9.52,5.58) {3};
\node[badge, draw=cterra, text=cterra] at (14.47,5.58) {4};
\node[badge, draw=cnavy, text=cnavy] at (19.42,5.58) {5};
\draw[arr] (4.02,2.7) -- (4.58,2.7) node[flow, midway] {$\bV$};
\draw[arr] (8.92,2.7) -- (9.48,2.7) node[flow, midway] {feat.};
\draw[arr, preaction={draw=white,line width=2.7pt}] (13.52,2.52) -- (14.43,2.52)
  node[flow, midway, fill=white, inner sep=0.7pt] {cond.};
\draw[arr, rounded corners=0.7mm] (atoms.east) -- (19.10,0 |- atoms.east) -- (19.10,0 |- cov.west) -- (cov.west);
\node[flow, fill=white, inner sep=1pt] at (19.10,2.70) {$\cA$};
\draw[arr] (23.42,2.7) -- (23.85,2.7) node[above=0.4mm, pos=0.8, font=\small] {$\hat\bH$};
\draw[dashed, cterra!85, line width=0.7pt, rounded corners=1.5mm]
  (trainloss.east) -- (24.25,0 |- trainloss.east) -- (24.25,5.94) -- (6.75,5.94);
\fill[cterra!85] (trainloss.east) circle (0.035);
\foreach \x in {6.75,11.50,16.625}
  \draw[-{Stealth[length=1.8mm]}, dashed, cterra!85, line width=0.7pt] (\x,5.94) -- (\x,5.62);
\node[font=\small, text=cterra!85, fill=white, inner sep=1pt] at (14.70,5.94)
  {loss gradients through fixed solve and covariance to stages 2--4};
\end{tikzpicture}}
\caption{The architecture of the TRACE estimator. Shared evidence and query information produce uncertain path atoms, which a fixed Bayesian solver combines with current pilots. Stages 2 to 4 are learning based.}
\label{fig:system}
\end{figure*}

\section{System model}
\label{sec:model}

We consider a single cell with single-antenna UEs and one BS. The BS at position $\bb \in \mathbb{R}^3$ is equipped with a half-wavelength spacing $N_1 \times N_2$ uniform planar array (UPA) with $N_{\mathrm r}=N_1N_2$ elements.
The receiver samples the channel at $K$ frequencies spaced by $\Delta f$ over bandwidth $B$, where sample $n$ has frequency offset $f_n=n\Delta f$ from the carrier $f_c$.
The BS observes each UE in short bursts spanning a few consecutive slots, during which the UE position is treated as constant and motion enters through the Doppler shift.
Each burst $i$ is transmitted by one UE. Since TRACE pools detections without using UE identity, we index all quantities by burst, and $q$ denotes the current burst of the query UE.
Path $l$ in burst $i$ has complex gain $\alpha_{il}$, delay $\tau_{il}$, Doppler shift $\nu_{il}$, azimuth $\phi_{il}$, and elevation $\theta_{il}$.
Its arrival direction at the BS is represented by the unit vector $\bu_{il}$, defined to point from the BS toward the incoming path.
Standard 5G New Radio (NR) receiver processing applies timing-advance control and frequency correction before pilot-based channel estimation~\cite{3gpp38213}. Since these operations are imperfect and UE-specific, the resulting CSI retains residual timing and frequency offsets and common complex distortions, so CSI from different UEs is not directly aligned in delay and Doppler. For burst $i$ and slot $t$, we model the observed CSI as~\cite{huang2026fuse}
\begin{equation}
\label{eq:channel}
\begin{split}
\widehat{\bH}_i[n,t] &= \beta_{i,t} \sum_{l} \alpha_{il}\,
e^{-j2\pi f_n (\tau_{il} + \delta_i)}\,
e^{j2\pi t T_s (\nu_{il} - \nu^{\mathrm{sync}}_i)} \\
&\quad\times \mathbf{a}(\phi_{il},\theta_{il}) + \mathbf{E}_i[n,t],
\end{split}
\end{equation}
where $\mathbf{a}(\phi,\theta) \in \mathbb{C}^{N_{\mathrm{r}}}$ is the array response with entries $e^{j2\pi\mathbf r_a^{\mathsf T}\bu}$, where $\bu$ is the unit vector of $(\phi,\theta)$ and $\mathbf r_a$ the position of element $a$ in wavelengths, and $T_s$ is the slot duration.
The scalar $\beta_{i,t}\in \mathbb{C}$ captures gain-control variation, oscillator phase noise, and residual common phase and carrier frequency offset (CFO) after phase tracking~\cite{wu2024bistaticasync}.
The residual timing offset (TO) $\delta_i$ after timing advance sums the UE clock offset and the estimation and quantization errors of the timing advance command, constant within a burst. The term $\nu^{\mathrm{sync}}_i$ is the common frequency absorbed by the coarse frequency synchronization loop.
The additive estimation noise is $\mathbf{E}_i[n,t]\sim\mathcal{CN}(\mathbf{0},\sigma_i^2\bI)$, independent across $n$ and $t$. The offsets are multiplicative and stay outside $\mathbf{E}_i[n,t]$.
Eq.~\eqref{eq:channel} holds for history and query bursts alike, but the two play different roles at the BS, as Table~\ref{tab:query_history} summarizes.

Few pilots leave the channel underdetermined, so past uplink traffic provides a prior on path directions and delays. Each PUSCH burst carries demodulation reference signals (DMRS) across its allocated band~\cite{3gpp38211}, which the BS already estimates for decoding, so this history requires no additional signaling.
A history burst $i$ keeps its own $\delta_i$, $\nu_i^{\mathrm{sync}}$, and noise variance $\sigma_i^2$. The BS stores with it the coarse position $\hbp_i$ logged during routine traffic, whose error scale $\sigma_p$ is a few meters. Sec.~\ref{ssec:sync} removes $\delta_i$ and $\nu_i^{\mathrm{sync}}$ and collects the result in a history set $\cD$.
Decoded PUSCH data can be remodulated into channel estimates for training.
For the current query burst $q$, the BS uses sparse SRS channel estimates and a coarse position estimate of the transmitting UE from 5G NR positioning~\cite{3gpp38305}.
SRS is shared among UEs through comb offsets and cyclic shifts, so we assume that only $\Np \in \{2,\dots,32\}$ of the $K$ frequency samples are available as pilots, indexed by a known set $\cP \subset \{0,\dots,K-1\}$. Each sample is a group of SRS subcarriers, so the $\Np$ pilots observe $\Np/K$ of the band.
The query is the special case $t=0$ of Eq.~\eqref{eq:channel}, so its Doppler and CFO factor equals one and the common phase sits in $\beta_q=\beta_{q,0}$. Our target $\bH\in\mathbb{C}^{K\times N_{\mathrm r}}$ is the noise-free part of that observation, $\bH[n]=\beta_q\sum_l \alpha_{ql}\,e^{-j2\pi f_n(\tau_{ql}+\delta_q)}\,\mathbf a(\phi_{ql},\theta_{ql})$.
The BS then observes $\by = [\bH]_{n \in \cP}+[\mathbf{E}_q[n,0]]_{n \in \cP} \in \mathbb{C}^{\Np N_{\mathrm{r}}}$, where $\sigma^2$ is the query noise variance, together with the coarse position $\hbp_q = \bp_q + \mathbf{e}_q$, $\mathbf{e}_q \sim \mathcal{N}(\mathbf{0}, \sigma_q^2 \bI)$. The served UE can be localized on demand with high-accuracy NR positioning, whereas historical positions are those logged during routine traffic, so we assume that $\sigma_q<\sigma_p$.
The simulations set $\delta_q=0$ and $\beta_q=1$.
\begin{table}[t]
\centering
\caption{History bursts, the query burst, and the target $\bH$.}
\label{tab:query_history}
\scriptsize
\setlength{\tabcolsep}{2pt}
\renewcommand{\arraystretch}{0.9}
\resizebox{\columnwidth}{!}{%
\begin{tabular}{@{}L{1.55cm}L{2.2cm}L{3.3cm}L{1.35cm}L{1.35cm}@{}}
\toprule
 & \textbf{History CSI} & \textbf{History detections} & \textbf{Query pilots} & \textbf{Target $\bH$} \\
\midrule
Symbol & $\widehat{\bH}_i[n,t]$ & $\Delta\tau_{ij}$, $\Delta\nu_{ij}$, $\mathbf{u}_{ij}$, $P_{ij}$ & $\by$ & $\bH$ \\
Receiver scalar & $\beta_{i,t}$, varies with slot & not applicable & $\beta_q$ & same $\beta_q$ \\
TO & $\delta_i$, per burst & cancelled by reference path & $\delta_q$ & same $\delta_q$ \\
Doppler/CFO & $\nu_{il}-\nu_i^{\mathrm{sync}}$ & cancelled by reference path & vanish $t=0$ & vanish $t=0$ \\
Noise & $\mathbf{E}_i[n,t]$ & reduced by detection & $\mathbf{E}_q[n,0]$ & none \\
Sync. & required & applied & not needed & not needed \\
Role & path extraction & voxel evidence & measurement & NMSE \\
\bottomrule
\end{tabular}}
\end{table}
Our goal is to find an estimator
$\hat\bH=\mathcal{G}(\by,\hbp_q,\Np,\sigma^2;\cD)$
that minimizes the NMSE, $\mathcal{G}_{\rm opt}=\arg\min_{\mathcal{G}}\frac{\E\|\hat\bH-\bH\|_F^2}{\E\|\bH\|_F^2}$.
Channel labels are normalized to unit mean power for each query, so this NMSE equals the NMSE averaged over queries, which Fig.~\ref{fig:main} and Table~\ref{tab:buckets} report.
Our inference uses no scene model, ray tracing, or stored CSI.
\section{TRACE: Learning Geometry-informed Channel Priors from Uplink Traffic}
\label{sec:method}

At upper mid-band, smooth building surfaces produce weak diffuse scattering~\cite{ying2025nyuray}, so we consider channels dominated by direct and reflected paths. Fig.~\ref{fig:system} shows the pipeline.

\subsection{Multi-UE Synchronization and History Construction}
\label{ssec:sync}
Detections from different UEs can only be fused if they share a common delay and Doppler reference. Each history UE has its own clock offset, timing-advance residual, and CFO, so we first synchronize each burst using its earliest strong arrival as the reference path~\cite{huang2026fuse}. The query UE skips this step, since it is not fused with other UEs and its common delay and phase belong to the target $\bH$.
Synchronizing a burst takes two steps. First, a sparse path detector (orthogonal matching pursuit with refinement) is run on the DMRS channel estimates of burst $i$. For each detected path $j$, a subset of the paths $l$, it gives the arrival direction $\bu_{ij}$, power $P_{ij}$, delay $\tau_{ij}$, and Doppler $\nu_{ij}$, all still shifted by $\delta_i$ and $\nu_i^{\mathrm{sync}}$.
Second, we subtract the delay $\tau_{i,\mathrm{ref}}$ and Doppler $\nu_{i,\mathrm{ref}}$ of the reference path and keep only the differences $\Delta\tau_{ij}=\tau_{ij}-\tau_{i,\mathrm{ref}}$ and $\Delta\nu_{ij}=\nu_{ij}-\nu_{i,\mathrm{ref}}$.
Because $\delta_i$ and $\nu_i^{\mathrm{sync}}$ are common to all paths within a burst, they cancel in these differences.
The referenced bursts form the history
$\cD=\{\hbp_i,\{\bu_{ij},\Delta\tau_{ij},\Delta\nu_{ij},P_{ij}\}_j\}_i$.

\suppressfloats[t]
\setcounter{figure}{1}
\begin{figure}[t]
\centering
\includegraphics[width=0.75\columnwidth]{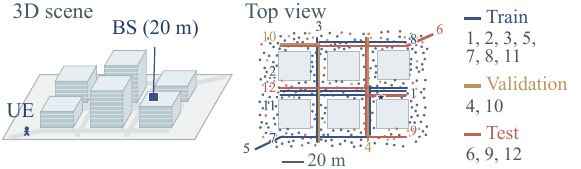}
\caption{Simulation scene with training, validation, test, and extra sampled positions. Overlapping routes are offset for visibility.}
\label{fig:scene}
\end{figure}
\setcounter{figure}{2}
\subsection{Voxel Evidence Representation}
\label{ssec:voxel}
To combine detections across UEs, we first map each detection into a common 3D space using its direction, relative delay, and coarse UE position.
For a detection at coarse UE position $\hbp$, we approximate the burst reference by the LoS path and estimate the path length as
$\ell^{\mathrm{path}}=\|\hbp-\bb\|_2+c\Delta\tau$, with $c$ the speed of light.
This estimate inherits UE position error and, for non-line-of-sight (NLoS) bursts, a common reference offset. We then project each detection in three ways. The first is a reflection point. Under a single-bounce model, it lies along $\bu$ from the BS, $\mathbf p^{\mathrm{refl}}=\bb+\ell^{\mathrm{BS}}\bu$. Its distance $\ell^{\mathrm{BS}}$ is fixed by the path length, $\ell^{\mathrm{BS}}+\|\hbp-\mathbf p^{\mathrm{refl}}\|_2=\ell^{\mathrm{path}}$, which with $\mathbf d=\hbp-\bb$ gives $\ell^{\mathrm{BS}}=\frac{(\ell^{\mathrm{path}})^2-\|\mathbf d\|^2}{2(\ell^{\mathrm{path}}-\bu^{\mathsf T}\mathbf d)}$.
The second is the corresponding image source,
$\mathbf p^{\mathrm{img}}=\hbp-\ell^{\mathrm{path}}
\frac{\hbp-\mathbf p^{\mathrm{refl}}}
{\|\hbp-\mathbf p^{\mathrm{refl}}\|_2}$.
The third is the simple projection
$\hbp-\ell^{\mathrm{path}}\bu$, which uses the BS-side direction directly. For a LoS path this point is the BS itself. For a reflection, it is not the image source and moves with the UE position, so we keep it only as a cheap directional cue and let the network decide its weight.
Since $\ell^{\mathrm{path}}\ge\|\mathbf d\|$, the denominator vanishes only for a LoS detection, where we replace it with a small constant and limit $\ell^{\mathrm{BS}}$ to the scene size, so every detection is kept.
These three projections, together with the historical UE position, form four types of spatial evidence. As their locations are uncertain, each contribution is distributed to the eight neighboring voxel centers with linearly decreasing weights.
Each geometric projection stores weighted detection count, log accumulated power, and burst support. The simple projection additionally stores relative-Doppler mean and spread and the position-error scale $\sigma_p$. These $C=13$ features form the evidence grid $\bV \in \mathbb{R}^{C \times N_{\mathrm h} \times N_{\mathrm h} \times N_{\mathrm v}}$ in Fig.~\ref{fig:system}, where $N_{\mathrm h}$ and $N_{\mathrm v}$ are the horizontal and vertical voxel counts of the BS-centered grid.

\setcounter{figure}{2}
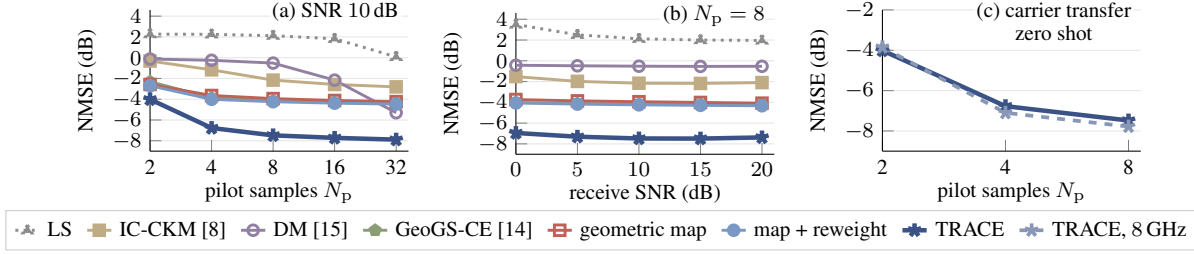
\begin{figure*}[t]
\centering
\begin{tikzpicture}
\begin{axis}[
  title={(a) SNR $10$\,dB}, title style={font=\footnotesize, at={(axis description cs:0.975,0.96)}, anchor=north east, inner sep=1pt, fill=white, fill opacity=0.8, text opacity=1},
  label style={font=\footnotesize}, tick label style={font=\footnotesize}, ymajorgrids, grid style={gray!20},
  enlarge x limits={upper, value=0.05}, clip marker paths=false,
  axis lines*=left, width=5.0cm, height=3.45cm, xlabel shift=-3pt,
  xlabel={pilot samples $\Np$}, ylabel={NMSE (dB)},
  xtick={2,4,8,16,32}, xticklabels={2,4,8,16,32}, xmode=log, log basis x=2,
  ymin=-9.0, ymax=4.6, ytick={4,2,0,-2,-4,-6,-8},
  legend to name=leg:main, legend columns=8,
  legend style={font=\footnotesize, draw=gray!40, column sep=0.5pt,
    /tikz/every even column/.append style={column sep=3pt}},
  legend image code/.code={\draw[mark repeat=2,mark phase=2,#1] plot coordinates {(0cm,0cm)(0.16cm,0cm)(0.32cm,0cm)};},
  legend cell align=left]
\addplot[black!45, dotted, line width=1.1pt, mark=triangle*, mark options={fill=black!45}] coordinates {(2,2.27)(4,2.25)(8,2.12)(16,1.82)(32,0.08)};
\addplot[msand, line width=1.1pt, mark=square*, mark options={fill=msand}] coordinates {(2,-0.32)(4,-1.17)(8,-2.16)(16,-2.58)(32,-2.81)};
\addplot[mpurple, line width=1.1pt, mark=o, mark options={fill=white}] coordinates {(2,-0.13)(4,-0.25)(8,-0.52)(16,-2.16)(32,-5.30)};
\addplot[mgreen, line width=1.1pt, mark=pentagon*, mark options={fill=mgreen}] coordinates {(2,-2.31)(4,-3.87)(8,-4.17)(16,-4.32)(32,-4.38)};
\addplot[mred, line width=1.1pt, mark=square, mark options={fill=white}] coordinates {(2,-2.58)(4,-3.64)(8,-3.94)(16,-4.12)(32,-4.23)};
\addplot[mblue, line width=1.1pt, mark=*, mark options={fill=mblue}] coordinates {(2,-2.70)(4,-4.01)(8,-4.24)(16,-4.38)(32,-4.47)};
\addplot[mnavy, line width=1.7pt, mark=star, mark size=2.6pt] coordinates {(2,-4.01)(4,-6.77)(8,-7.48)(16,-7.72)(32,-7.89)};
\addlegendimage{mnavy!60, dashed, line width=1.4pt, mark=star, mark size=2.6pt, mark options={solid}}

\legend{LS, IC-CKM \cite{icckm2025}, DM \cite{zhou2025diffusion}, GeoGS-CE \cite{geogs2026}, geometric map, map + reweight, TRACE, {TRACE, $8$\,GHz}}
\end{axis}
\end{tikzpicture}\hspace{2mm}
\begin{tikzpicture}
\begin{axis}[
  title={(b) $\Np = 8$}, title style={font=\footnotesize, at={(axis description cs:0.975,0.96)}, anchor=north east, inner sep=1pt, fill=white, fill opacity=0.8, text opacity=1},
  label style={font=\footnotesize}, tick label style={font=\footnotesize}, ymajorgrids, grid style={gray!20},
  enlarge x limits={upper, value=0.05}, clip marker paths=false,
  axis lines*=left, width=5.0cm, height=3.45cm, xlabel shift=-3pt,
  xlabel={receive SNR (dB)}, ylabel={NMSE (dB)},
  xtick={0,5,10,15,20}, ymin=-9.0, ymax=4.6, ytick={4,2,0,-2,-4,-6,-8}]
\addplot[black!45, dotted, line width=1.1pt, mark=triangle*, mark options={fill=black!45}] coordinates {(0,3.50)(5,2.50)(10,2.12)(15,2.00)(20,1.96)};
\addplot[msand, line width=1.1pt, mark=square*, mark options={fill=msand}] coordinates {(0,-1.52)(5,-1.98)(10,-2.16)(15,-2.18)(20,-2.10)};
\addplot[mpurple, line width=1.1pt, mark=o, mark options={fill=white}] coordinates {(0,-0.43)(5,-0.48)(10,-0.52)(15,-0.54)(20,-0.53)};
\addplot[mgreen, line width=1.1pt, mark=pentagon*, mark options={fill=mgreen}] coordinates {(0,-4.02)(5,-4.11)(10,-4.17)(15,-4.19)(20,-4.18)};
\addplot[mred, line width=1.1pt, mark=square, mark options={fill=white}] coordinates {(0,-3.73)(5,-3.86)(10,-3.94)(15,-4.01)(20,-4.08)};
\addplot[mblue, line width=1.1pt, mark=*, mark options={fill=mblue}] coordinates {(0,-4.06)(5,-4.17)(10,-4.24)(15,-4.29)(20,-4.33)};
\addplot[mnavy, line width=1.7pt, mark=star, mark size=2.6pt] coordinates {(0,-6.95)(5,-7.31)(10,-7.48)(15,-7.49)(20,-7.38)};
\end{axis}
\end{tikzpicture}\hspace{2mm}
\begin{tikzpicture}
\begin{axis}[
  title={(c) carrier transfer\\ zero shot}, title style={font=\footnotesize, at={(axis description cs:0.975,0.96)}, anchor=north east, align=center, inner sep=1pt, fill=white, fill opacity=0.8, text opacity=1},
  label style={font=\footnotesize}, tick label style={font=\footnotesize}, ymajorgrids, grid style={gray!20},
  enlarge x limits={upper, value=0.05}, clip marker paths=false,
  axis lines*=left, width=5.0cm, height=3.45cm, xlabel shift=-3pt,
  xlabel={pilot samples $\Np$}, ylabel={NMSE (dB)},
  xmode=log, log basis x=2, xtick={2,4,8}, xticklabels={2,4,8},
  ymin=-9.0, ymax=-2.0, ytick={-2,-4,-6,-8},
]
\addplot[mnavy, line width=1.7pt, mark=star, mark size=2.6pt] coordinates {(2,-4.01)(4,-6.77)(8,-7.48)};
\addplot[mnavy!60, dashed, line width=1.4pt, mark=star, mark size=2.6pt, mark options={solid}] coordinates {(2,-3.85)(4,-7.10)(8,-7.78)};
\end{axis}
\end{tikzpicture}\\[-0.5mm]
\pgfplotslegendfromname{leg:main}
\caption{Channel estimation NMSE of the TRACE network trained at $15$\,GHz, $10$\,dB, and $\Np \in \{2,4,8\}$, evaluated along three deployment axes: (a) varying pilot count at $10$\,dB, (b) varying SNR at $\Np=8$, and (c) zero shot carrier transfer to $8$\,GHz without adaptation.}
\label{fig:main}
\end{figure*}

\subsection{Query-Conditioned Path Reconstruction}
\label{ssec:unet}
UEs share the environment but follow different paths.
We therefore encode the voxel evidence grid $\bV$ once and condition it on each query.
Fixed Fourier features and a small neural network turn the coarse position, pilot count, noise variance, and the signal power estimated from the pilots into a query vector $\mathbf q$.
FiLM~\cite{perez2018film} scales and shifts the features by $\mathbf q$ so the network can adjust how strongly it relies on the prior, self-attention connects spatially distant evidence, and cross-attention selects the evidence relevant to the query UE.
The decoder scores all voxels and retains the $M=48$ highest scoring locations. For each selected voxel $m$, a bounded 3D offset refines its center to $\mathbf p_m$. The decoder also predicts delay and angular uncertainty widths $\gamma_m^\tau$ and $\gamma_m^{\mathrm a}$, three geometry scores $s_m^{(k)}$, and a power score. The power prediction is initialized from the image-source count feature.
Each selected location $\mathbf p_m$ is read by three geometric branches, as a scatterer between the BS and UE, as a mirrored BS, or as the direct path. Each branch produces a candidate delay and arrival direction $(\tau_m^{(k)},\bu_m^{(k)})$, where
$\mathcal B=\{\mathrm{scat},\mathrm{mir},\mathrm{dir}\}$.
The branch scores are converted to mixing weights
\begin{equation}
\label{eq:branchmix}
\begin{aligned}
\pi_m^{(k)}&=\frac{\exp(s_m^{(k)})}
{\sum_{r\in\mathcal B}\exp(s_m^{(r)})},
&\tau_m&=\sum_{k\in\mathcal B}\pi_m^{(k)}\tau_m^{(k)},\\
\widetilde{\bu}_m&=\sum_{k\in\mathcal B}\pi_m^{(k)}\bu_m^{(k)},
&\bu_m&=\frac{\widetilde{\bu}_m}{\|\widetilde{\bu}_m\|_2}.
\end{aligned}
\end{equation}
The mirrored-BS branch is given the largest initial score, while all branch scores are learned through the channel-estimation loss.
We additionally keep an explicit LoS atom at $m=0$. Its direction and delay follow from the coarse query position, while its weight and uncertainty widths come from $\mathbf q$. Together with the selected voxel outputs, these form
$\cA=\{(\bu_m,\tau_m,\gamma_m^\tau,\gamma_m^{\mathrm a},w_m)\}_{m=0}^{M}$.
The weights are normalized so that their sum matches the estimated signal power, computed from the mean pilot power after subtracting $\sigma^2$.
Only this power scale changes with the current pilots, while the atom directions, delays, widths, and query vector are computed once and reused across slots, so the estimator stays linear in $\by$ given the prior.%

\subsection{Covariance-Based Channel Estimation}
\label{ssec:solve}
The atom set $\cA$ describes the likely path structure but leaves the current complex path gains unknown. We therefore use it to construct a covariance $\bR$ linking observed pilots to unobserved frequencies.
For each atom, let $\mathbf D_m\in\mathbb C^{K\times K}$ and $\mathbf A_m\in\mathbb C^{N_{\mathrm r}\times N_{\mathrm r}}$ describe the frequency and antenna correlation,
\begin{equation}
\label{eq:kernels}
\begin{aligned}
[\mathbf D_m]_{n,n'} &= e^{-j2\pi(f_n-f_{n'})\tau_m}\,
\operatorname{sinc}\!\big((f_n-f_{n'})\gamma_m^\tau\big),\\
[\mathbf A_m]_{a,b} &= e^{j2\pi\mathbf d_{ab}^{\mathsf T}\bu_m}\,
\operatorname{sinc}\!\big(\gamma_m^{\mathrm a}\|(\bI-\bu_m\bu_m^{\mathsf T})\mathbf d_{ab}\|\big),
\end{aligned}
\end{equation}
where $n,n'$ index frequencies, $a,b$ index antennas, $\mathbf d_{ab}=\mathbf r_a-\mathbf r_b$ is their displacement in carrier wavelengths, and $\operatorname{sinc}(x)=\sin(\pi x)/(\pi x)$.
The exponential terms are the covariance of an exact plane wave at $\tau_m$ and $\bu_m$, and the sinc terms encode the error these inherit from the coarse position and the voxel.
Averaging the delay over a spread $\gamma_m^\tau$ gives the delay sinc, and the angular sinc is a separable sinc approximation of the same effect across antennas.
A zero width ($\operatorname{sinc}=1$) leaves an exact plane wave and a large width leaves samples nearly uncorrelated, so the widths say how far to trust each atom.
Assuming zero-mean and mutually uncorrelated atom gains, we have
$\bR = \sum_{m=0}^{M} w_m \big(\mathbf D_m\otimes\mathbf A_m\big)\in\mathbb C^{KN_{\mathrm r}\times KN_{\mathrm r}}$, with $\otimes$ the Kronecker product. Neighboring atoms on the same physical path share one gain, but their overlapping kernels add up like a single atom with larger widths, which the learned widths account for.
Stack antennas within each frequency as $\bh=\operatorname{vec}(\bH^{\mathsf T})\in\mathbb C^{KN_{\mathrm r}}$.
The pilot set $\cP$ selects the pilot covariance $\bR_{\mathrm{pp}}\in\mathbb C^{\Np N_{\mathrm r}\times\Np N_{\mathrm r}}$ and the cross covariance $\bR_{\mathrm{fp}}\in\mathbb C^{KN_{\mathrm r}\times\Np N_{\mathrm r}}$ between the full channel and the pilots.
Using $\bR$ as a prior covariance gives the LMMSE form
\begin{equation}
\label{eq:lmmse}
\hat\bh=\bR_{\mathrm{fp}}(\bR_{\mathrm{pp}}+\sigma^2\bI)^{-1}\by,
\end{equation}
which is reshaped into $\hat\bH$. We train through the fixed solver by minimizing mean $\log(\|\hat\bH-\bH\|_F^2/\|\bH\|_F^2)$ over queries, with the noise-free channels $\bH$ as targets.

\section{Simulation Results}
\label{sec:results}

\begin{table}[b]
\centering
\renewcommand{\arraystretch}{0.87}
\caption{Scenario, network, and training parameters.}
\label{tab:setup}
\fontsize{7}{7.5}\selectfont
\setlength{\tabcolsep}{2pt}
\resizebox{\columnwidth}{!}{%
\begin{tabular}{@{}ll@{\hspace{6pt}}ll@{\hspace{6pt}}ll@{}}
\toprule
Carrier $f_c$ & $15$\,GHz & Bandwidth $B$ & $200$\,MHz & Samples $K$, $\Delta f$ & 64, $3.125$\,MHz \\
BS array & $8\times8$, $\lambda/2$ & Scene positions & 24\,254 & Traced paths & 582\,410 \\
History bursts (det.) & 1\,259 (9\,963) & History $\sigma_p$ & $3$\,m & Query $\sigma_q$ & $0.5$\,m \\
Grid $N_{\mathrm{h}}$, $N_{\mathrm{v}}$, $C$ & 192, 16, 13 & Spacing (h, v) & $1.53$, $5.81$\,m & Atoms $M{+}1$ & 49 \\
Parameters & 15.3M & Optimizer & AdamW & Peak learning rate & $3\times10^{-4}$ \\
Steps, batch & 8\,000, 8 & Schedule & warmup, constant & Precision & bf16 \\
Training SNR & $10$\,dB & Queries (val, test) & 492, 837 & Test LoS fraction & $49\%$ \\
\bottomrule
\end{tabular}}
\end{table}

\begin{table}[t]
\centering
\caption{NMSE (dB) at $\Np=8$, $10$\,dB. Map uses the same traffic without learning, so the gap to TRACE is what learning adds.}
\label{tab:buckets}
\fontsize{7}{7.5}\selectfont
\setlength{\tabcolsep}{2pt}
\renewcommand{\arraystretch}{0.87}
\resizebox{\columnwidth}{!}{%
\begin{tabular}{@{}lrr@{\hspace{10pt}}lrr@{}}
\toprule
\textbf{Condition} & \textbf{TRACE} & \textbf{Map} & \textbf{Change} & \textbf{TRACE} & \textbf{Map} \\
\midrule
All (837) & $-7.48$ & $-3.94$ & $60\%$ fewer history bursts & $-7.27$ & $-3.94$ \\
LoS (407) & $-13.02$ & $-5.31$ & Position error $\sigma_q=0$\,m & $-7.53$ & $-4.08$ \\
NLoS (430) & $-5.22$ & $-2.96$ & Position error $\sigma_q=1$\,m & $-6.99$ & $-3.46$ \\
No anchor (156) & $-4.97$ & $-3.23$ & Position error $\sigma_q=3$\,m & $-3.23$ & $-2.07$ \\
Oracle atoms, $\hbp_q$ & $-13.29$ & -- & Diffraction, scattering on & $-6.77$ & $-4.21$ \\
Oracle atoms, $\bp_q$ & $-20.19$ & -- & Plus map anchors & $-6.94$ & -- \\
\bottomrule
\end{tabular}}
\end{table}

We evaluate TRACE in the \texttt{simple\_street\_canyon} scene of Sionna RT~\cite{sionna}, traced with up to five reflections. The main results keep diffraction and diffuse scattering disabled, and Sec.~\ref{ssec:robust} repeats the test with both enabled.
Fig.~\ref{fig:scene} shows the scene and route split. Table~\ref{tab:setup} the settings.
Each history burst is rendered from its ray-traced paths with its own offsets. Following 5G NR~\cite{3gpp38213}, $\delta_i\sim\mathcal U[0,65]$\,ns\ ($65$\,ns matches the NR timing-advance step for $\mu=3$) and $\nu_i^{\mathrm{sync}}$ follows a slow random walk within $\pm25$\,Hz. The burst is corrupted by noise at $15$\,dB and passed to the detector of Sec.~\ref{ssec:sync}, which returns $7.9$ paths on average, so detection errors come from the detector itself.
With 2 to 8 pilots, only $3\%$ to $12.5\%$ of the 64 frequency samples are observed. The pilots are spread evenly over the band with no random offset, and the same set $\cP$ is used for every query.
For each query, $\sigma^2$ is set from the channel power to achieve the target SNR.
Standard timing and phase tracking keep the query offsets $\delta_q$ and $\beta_q$ small and they are not our subject, so we set $\delta_q=0$ and $\beta_q=1$.
The target is then the ray-traced channel with no TO and, since $t=0$, no CFO.
We train one network at $15$\,GHz, $10$\,dB, and $\Np\in\{2,4,8\}$, sampling one $\Np$ for each batch, and reuse it in every test unless stated.
We split UE walking routes across training, validation, and test, with some test segments unseen in training.
Inference uses no true positions, paths, or query channels.
On one RTX 4090, encoding $\mathbf{V}$ takes $13.6$\,ms once, and each query adds about $20$\,ms batched or $26$\,ms alone.

\subsection{Baseline Comparison}
\label{ssec:comparison}

Fig.~\ref{fig:main} compares TRACE with least squares (LS) interpolation, three SOTA priors, and two geometric ablations.
IC-CKM~\cite{icckm2025} is favored in two ways. Its map uses the clean full-band CSI of the history bursts, pooled from nearby cells, where TRACE only gets noisy detections, and its solver inflates $\sigma^2$ by $\nu=30$, tuned on validation, so that map error counts as noise.
The diffusion model (DM)~\cite{zhou2025diffusion} is trained on our 5422 clean training channels, and samples 100 reverse steps in the delay-angle domain with pilots entering through a likelihood gradient. GeoGS-CE~\cite{geogs2026} is adapted to uplink queries with its explicit LoS component.
The geometric map generates candidate paths and weights from deterministic geometry.
Map with reweighting uses the same candidate paths and learns their weights and uncertainty scales.
TRACE instead learns the candidate locations, their confidence, and their widths.
All methods share the test queries and pilots, and the last three share the solver.
At 8 pilots, TRACE reaches $-7.48\pm0.01$\,dB (over 6 noise realizations).
LS stays above $0$\,dB because with pilots $25$\,MHz apart, delays beyond $40$\,ns alias.
IC-CKM reaches $-2.16$\,dB. Only $44\%$ of test queries have history in their own $5$\,m cell, so it borrows CSI from nearby cells, which improves coverage but introduces position mismatch.
DM reaches $-0.5$\,dB NMSE at 8 pilots, the NMSE of keeping the pilots and zeroing the other $56$ samples. Its prior knows neither query position nor history, so it returns its mean for unobserved samples. A stronger sampler and a wider denoiser only change this by under $0.01$\,dB.
GeoGS-CE reaches $-4.17$\,dB, mostly from its explicit LoS modeling.
TRACE outperforms the geometric map by $3.13$ and $3.53$\,dB on average at 4 and 8 pilots.
Reweighting the fixed candidates accounts for only $0.30$\,dB of the gain at 8 pilots. The other $3.23$\,dB comes from learning candidate geometry and path confidence.
At 2 pilots, TRACE still outperforms the strongest baseline by $1.31$\,dB.
Up to 32 pilots, TRACE keeps improving despite both map curves flatten because their candidates are fixed. DM becomes the strongest baseline once the pilots carry enough delay structure, still $2.59$\,dB behind TRACE.

\subsection{Analysis of Performance Gains}
\label{ssec:anatomy}

Table~\ref{tab:buckets} splits the gain over the geometric map by query condition. A query has no anchor when the BS cannot reach it by a direct or single-bounce path. The geometric map uses 8 fixed candidate paths: one LoS path and seven image-source or scatterer paths derived from training detections.
The gain is larger in LoS, $7.7$\,dB versus $2.3$\,dB in NLoS.
A LoS channel is dominated by one path, and the learned atoms recover its direction and delay. Queries without an anchor gain the least, so TRACE helps less where no single-bounce path exists, which the atom branches can only approximate.
Two ablations show where the gain comes from. Without the image-source count that initializes the power score, TRACE falls to $-3.91$\,dB, the level of the geometric map, so the learned refinement builds on the geometric anchoring rather than replacing it.
Without the simple projection, TRACE loses $0.6$\,dB on average over three seeds at 4 and 8 pilots, mainly because only this projection carries the Doppler and $\sigma_p$ features.
Adding the map anchors to the learned atoms makes NMSE worse, so the learned atoms already cover them. Oracle atoms from each query's own ray-traced paths, solved with the same LMMSE estimator, show what perfect geometry could give. At the same coarse position, TRACE recovers $56\%$ of their gain in dB.
With the true position instead of the coarse one ($\sigma_q=0.5$\,m), the oracle improves further, so better UE localization would help further.

\subsection{Deployment Robustness}
\label{ssec:robust}

To test SNR robustness, we apply TRACE trained at $10$\,dB directly from $0$ to $20$\,dB without retraining. It outperforms all baselines at every SNR, while its NMSE varies by only $0.54$\,dB.
For carrier transfer, we retrace the scene at $8$\,GHz with the same bandwidth, samples, and electrical array, rebuild the history and voxel grid from the $8$\,GHz traffic, and reuse the network. TRACE stays within $0.35$\,dB of its $15$\,GHz results, since the geometric atoms reproject the prior to the new carrier.
Removing $60\%$ of the history bursts before building the voxel grid costs only $0.21$\,dB at 8 pilots over five random subsets.
Trained only at $\sigma_q=0.5$\,m, TRACE loses little at $1$\,m, but at $3$\,m the coarse position no longer points at the right specular paths and every position-based estimator collapses.
Retracing the scene with diffraction and diffuse scattering (coefficient $0.1$) multiplies the paths by about forty while specular paths keep $99.6\%$ of the power, and costs $0.71$\,dB.
We test scene transfer on four unseen scenes, two street canyons of different width and parts of the Munich and Paris \'Etoile scenes of Sionna RT, which come from OpenStreetMap data~\cite{openstreetmap}.
Zero shot NMSE is $-1.7$, $-2.7$, $-0.8$, and $-3.1$\,dB, while fine-tuning on 400 target scene queries gives $-2.4$, $-5.3$, $-8.1$, and $-5.4$\,dB.

\section{Conclusion}
\label{sec:conclusion}

We proposed TRACE to learn uncertain path hypotheses from routine uplink detections and combine them with current pilots through a fixed solver. With 8 pilots, it outperformed three SOTA priors and two geometric ablations by more than $3$\,dB and stayed robust to pilot count, SNR, carrier frequency, and reduced mapping traffic. Future work will extend TRACE to dynamic propagation environments.
\vfill\pagebreak

\section{Acknowledgment}
\label{sec:ack}
This work was supported in part by the Industry Affiliate Program, Center for Wireless Communications (CWC), University of California San Diego.
The authors used Claude (Anthropic) to improve the grammar, clarity, and readability of the text in all sections. The AI tool was not used to generate technical content or results.

\section{Compliance with Ethical Standards}
As this study involved only numerical simulations and no human participants or animal subjects, ethical approval was not required.

\bibliographystyle{IEEEbib}
\bibliography{refs}

\end{document}